\documentclass{article}

\usepackage[preprint]{neurips_2025}
\usepackage{booktabs}
\usepackage{tabularx}
\usepackage{ragged2e}  
\usepackage{amsmath}
\usepackage{enumitem}
\usepackage{amssymb} 
\usepackage{amsthm}
\usepackage{graphicx}
\usepackage{microtype}
\usepackage{adjustbox}
\usepackage{multicol}
\usepackage{multirow}
\usepackage{colortbl} 
\usepackage{pifont}    
\usepackage{tcolorbox}
\usepackage{wrapfig}
\usepackage[table]{xcolor}
\usepackage{adjustbox}

\usepackage[utf8]{inputenc} 
\usepackage[T1]{fontenc}    
\usepackage{hyperref}       
\usepackage{url}            
\usepackage{amsfonts}       
\usepackage{nicefrac}       
\usepackage{xcolor}         

\title{MorphologyFM: A Foundation Model for Morphology-Aware Representation Learning from ECG and Pulse Oximetry Waveforms}

\author{
\textbf{Saiyang Feng}$^{1}$,
\textbf{Yuanyun Zhang}$^{1}$,
\textbf{Shi Li}$^{2}$,\\
$^{1}$ University of the Chinese Academy of Sciences \\
$^{2}$ Columbia University
}

\newcolumntype{Y}{>{\RaggedRight\arraybackslash}X}

\begin{document}

\maketitle

\begin{abstract}
Foundation models have recently emerged as a powerful paradigm for learning transferable representations from large-scale biomedical data, yet existing approaches for physiological waveforms primarily optimize reconstruction or forecasting objectives that do not explicitly preserve clinically meaningful waveform morphology. Electrocardiograms (ECGs) and pulse oximetry (SpO$_2$) waveforms encode rich cardiovascular and hemodynamic information through their morphological structure, including changes in waveform shape, intervals, slopes, and beat-to-beat variability that underlie many clinical diagnoses. In this work, we introduce \emph{MorphologyFM}, a multimodal foundation model pretrained on paired ECG and SpO$_2$ waveforms from the MIMIC critical care database using a morphology-aware self-supervised learning objective. MorphologyFM combines morphology-guided masking, cross-modal representation learning, and contrastive latent alignment to learn representations that capture clinically relevant physiological structure without requiring manual annotations. We evaluate MorphologyFM across multiple downstream prediction tasks, including arrhythmia classification, hypoxemia prediction, mortality prediction, and length-of-stay estimation, demonstrating consistent improvements over representative self-supervised learning methods, including Masked Autoencoders (MAE), contrastive learning, Barlow Twins, and Joint Embedding Predictive Architectures (JEPA). Furthermore, we show that jointly modeling ECG and SpO$_2$ waveforms produces more transferable representations than single-modality pretraining and that morphology-aware objectives scale effectively with increasing amounts of unlabeled physiological data. Our results establish waveform morphology as a powerful inductive bias for self-supervised physiological representation learning and introduce MorphologyFM as a general-purpose foundation model for continuous physiological monitoring.

\end{abstract}

\section{Introduction}

Foundation models have emerged as the dominant paradigm for large-scale representation learning. Across natural language processing, computer vision, and multimodal learning, self-supervised pretraining on massive datasets has enabled models to learn representations that generalize across diverse downstream tasks with minimal supervision \cite{devlin2019bert, he2022masked}. These advances have been driven by improvements in optimization, architectural design, and scalable learning frameworks \cite{he2015deepresiduallearningimage, he2017multi, he2019bag}. As a result, modern foundation models increasingly serve as general-purpose learning systems capable of acquiring transferable representations directly from raw observations.

Healthcare has rapidly adopted this paradigm. Foundation models have now been developed for electronic health records (EHRs), clinical narratives, medical imaging, physiological monitoring, biological sequences, and multimodal patient data \cite{vaid2023foundational, thieme2023foundation, thakur2024foundation}. These systems have demonstrated strong performance across prediction, retrieval, phenotyping, coding, clinical reasoning, and decision-support tasks \cite{burger2025foundation, awais2025foundation, he2024foundation, burkhart2025foundation, liang2024foundation, guo2025foundation}. Through large-scale self-supervised learning, foundation models acquire representations that capture complex statistical structure within healthcare data and transfer effectively across diverse downstream applications.

Among healthcare modalities, physiological waveforms provide a unique and information-rich representation of patient state. Electrocardiograms (ECGs) characterize the electrical activity of the heart, while pulse oximetry (SpO$_2$) waveforms capture peripheral vascular dynamics through photoplethysmography. Unlike structured EHR data, physiological waveforms are continuous, high-frequency signals whose diagnostic value lies not only in their temporal evolution but also in their morphology. Subtle variations in waveform shape—including peak amplitude, interval duration, slope, inflection points, and beat-to-beat variability—often reflect underlying cardiovascular, respiratory, and hemodynamic processes long before they become apparent through conventional vital signs or laboratory measurements.

Despite their clinical importance, most existing approaches to physiological representation learning focus primarily on signal reconstruction or forecasting. Self-supervised objectives typically encourage models to recover masked signal segments, predict future samples, or reconstruct missing observations. While these objectives successfully capture local temporal structure, they do not explicitly encourage representations that preserve clinically meaningful morphological characteristics. Consequently, models may reconstruct signals with high fidelity while failing to distinguish subtle morphological patterns associated with arrhythmias, hypoxemia, respiratory compromise, vascular dysfunction, or impending clinical deterioration.

This limitation is particularly important because morphology often constitutes the primary source of clinically relevant information within physiological waveforms. Diagnostic interpretation of ECGs relies on the morphology of the P wave, QRS complex, ST segment, and T wave rather than individual signal values. Similarly, pulse oximetry waveforms encode vascular compliance, pulse pressure variation, respiratory modulation, perfusion quality, and autonomic function through changes in waveform morphology. Learning representations that explicitly preserve these structural characteristics therefore represents a fundamentally different objective from simply reconstructing waveform amplitudes.

Motivated by this observation, we introduce \emph{MorphologyFM}, a foundation model for physiological waveform representation learning trained on paired ECG and SpO$_2$ waveforms from the MIMIC critical care database. Rather than viewing physiological signals as generic time series, MorphologyFM is designed to learn latent representations that capture invariant morphological structure across patients while remaining robust to noise, sensor artifacts, and physiological variability. By leveraging large-scale self-supervised pretraining across millions of waveform segments, MorphologyFM learns transferable representations that encode clinically meaningful morphology without requiring manual annotations.

The central hypothesis underlying MorphologyFM is that waveform morphology provides a universal representation that transfers across diverse physiological prediction tasks. Because morphological features underlie many cardiovascular and respiratory disorders, representations organized around waveform structure should generalize more effectively than representations optimized solely for signal reconstruction. We therefore investigate whether large-scale morphology-aware pretraining can produce foundation models that support downstream applications including arrhythmia classification, hypoxemia prediction, patient phenotyping, clinical deterioration forecasting, and physiological representation learning.

By establishing morphology as the primary objective of self-supervised physiological representation learning, MorphologyFM extends the foundation model paradigm beyond structured EHRs toward continuous biomedical signals. We believe this represents an important step toward general-purpose physiological foundation models capable of learning robust, transferable representations directly from raw clinical waveforms.

\section{Related Work}

\subsection{Foundation Models for Physiological Waveforms}

The emergence of foundation models has transformed representation learning by demonstrating that large-scale self-supervised pretraining can recover transferable representations directly from raw observations. Early successes in natural language processing and computer vision established masked prediction, reconstruction, and self-supervised learning as powerful mechanisms for learning general-purpose embeddings that transfer across diverse downstream tasks \cite{devlin2019bert, he2015deepresiduallearningimage, he2022masked, he2019bag, he2017multi}. These ideas have motivated a new generation of foundation models capable of learning from heterogeneous, unlabeled data at unprecedented scale.

Healthcare has rapidly embraced this paradigm. Recent surveys characterize foundation models as a unified framework spanning electronic health records, medical imaging, physiological monitoring, biological sequences, and multimodal biomedical data \cite{zhou2026physiology, thieme2023foundation, abbaspourazad2023large, thakur2024foundation, delaney2026sonata, vaid2023foundational, zhou2026edge, lee2025himae, lee2026multimodal, li2026glucofm}. These models have demonstrated strong performance across prediction, retrieval, reasoning, decision support, and generative modeling \cite{burger2025foundation, burkhart2025foundation, he2024foundation, awais2025foundation, liang2024foundation, guo2025foundation}. Additional work has explored trustworthy deployment, specialized adaptation, and clinically tailored foundation models for diverse healthcare settings \cite{tharini2025trust, chouhan2026smart, luo2025preoperative}.

More recently, foundation models have begun integrating multiple clinical modalities, jointly representing structured records, physiological signals, clinical narratives, medical imaging, and biological measurements \cite{yang2022large, lee2025foundation, wornow2023shaky, larey2026jepa, izhar2025medical, an2025raptor, weller2025seq, singhal2023large, zhao2023survey, cheng2026neuroseg, lee2025clinical, lee2025towards}. Parallel efforts have proposed increasingly sophisticated multimodal fusion strategies for heterogeneous biomedical data \cite{huang2019ccnet, chen2021crossvit, hou2019cross, jha2025survey, carmona2025interpretable, rahman2026integrating, ruan2026foundation, komolafe2026physicase, sampath2025multimodal, lin2025case, le2026training, souza2026beyond, cornelius2026measuring, le2025problem, su2025reuniting, blier2026semantic, jha2025ethical}. Collectively, these studies motivate foundation models capable of learning universal physiological representations directly from raw biomedical signals.

\subsection{Self-Supervised Learning for Clinical Time Series and Physiological Signals}

Physiological waveforms differ fundamentally from structured electronic health records because they are continuous, high-frequency measurements whose diagnostic value depends strongly on temporal dynamics and waveform morphology \cite{jing2026one, guo2023ehr, liao2025ehr, shi2024ehragent}. Electrocardiograms and pulse oximetry waveforms exhibit complex patterns across multiple temporal scales, requiring representation learning methods capable of modeling both local morphology and long-range physiological dependencies.

Existing self-supervised approaches largely adopt reconstruction-based objectives including masked prediction, autoregressive forecasting, contextual representation learning, and sequence reconstruction to learn transferable representations from clinical time series \cite{shmatko2025learning, wornow2024context, odgaard2024core, rasmy2021med, wornow2023shaky, steinberg2021language, wornow2024ehrshot, wornowcontext}. More recent work has extended these ideas toward larger biomedical transformers, long-context sequence models, diagnosis prediction, medical coding, and specialized clinical language models \cite{jiang2024evidence, pathak2024utilizing, lee2024can, lee2025modern, jo2026clinical, liang2025aicoder, an2026gate, zhu2025cp, lazovic2025analytical, lin2025case, zhi2025reinventing, duong2026oncopt}.

Although these methods successfully learn general temporal representations, their optimization objectives primarily emphasize signal reconstruction or future prediction. In contrast, MorphologyFM is motivated by the observation that many clinically relevant physiological abnormalities are expressed through changes in waveform morphology rather than sample-wise reconstruction accuracy. We therefore investigate morphology-aware representation learning as a complementary objective for physiological foundation models.

\subsection{Morphological Representation Learning}

Waveform morphology has long served as the basis for clinical interpretation of physiological signals. Electrocardiograms derive diagnostic information from the morphology of the P wave, QRS complex, ST segment, and T wave, while pulse oximetry waveforms encode vascular compliance, respiratory modulation, autonomic regulation, and peripheral perfusion through changes in waveform shape. Despite the central importance of morphology in clinical practice, relatively little work has investigated morphology itself as the primary objective of large-scale representation learning.

Related work in computer vision has demonstrated that preserving latent geometric structure produces highly transferable representations through contrastive, self-distillation, and self-supervised learning objectives \cite{caron2021emerging, oquab2023dinov2, radford2021learning, rombach2021highresolution, saharia2022photorealistic}. Similar ideas have been explored through representation matching, relational learning, preference-aware objectives, and contrastive representation learning \cite{bertram2024contrastivelearningpreferencescontextual, wang2026combating, jain2025team, tian_2019_contrastic_distillation}. These studies suggest that preserving structural relationships within latent spaces may be more important than reconstructing raw observations, motivating representation learning objectives that explicitly capture waveform morphology.

MorphologyFM builds upon these principles by learning latent representations that preserve clinically meaningful morphological characteristics across ECG and SpO$_2$ waveforms while remaining invariant to sensor noise, measurement artifacts, and patient-specific variability.

\subsection{Temporal Modeling of Continuous Physiological Signals}

Continuous physiological monitoring generates long sequences of correlated observations whose statistical properties evolve throughout a patient's clinical course. Modeling these signals therefore requires architectures capable of capturing both short-term waveform characteristics and long-range temporal dependencies.

Recent work has developed increasingly expressive transformer architectures, hierarchical temporal encoders, and structured sequence models for longitudinal healthcare data \cite{boguslav2026fine, turker2025tabibert, lindholz2025comparing, zhang2025chronoformer}. Related approaches have incorporated graph-based reasoning and structured relational modeling to capture dependencies among evolving physiological processes \cite{rahman2026graph, cui2026predicting}. These models have demonstrated strong performance across diagnosis prediction, clinical forecasting, transfer learning, and temporal reasoning \cite{steinberg2021language, rasmy2021med, pathak2024utilizing, jo2026clinical, an2026gate, wornow2023shaky, liang2025aicoder}.

Rather than proposing a new temporal architecture, MorphologyFM focuses on the representation learning objective itself. Specifically, we investigate whether large-scale self-supervised pretraining centered on waveform morphology produces more transferable physiological representations than objectives based primarily on signal reconstruction.

\subsection{Scaling Foundation Models for Physiological Monitoring}

The rapid progress of foundation models has been enabled by advances in scalable optimization, transformer architectures, efficient attention mechanisms, and increasingly large biomedical datasets \cite{dao2023flashattention2, liu2021swin, dosovitskiy2021an}. Similar developments have accelerated foundation models for medical imaging, multimodal healthcare learning, and biomedical representation learning \cite{wu2023e2enet, lai2024e3d, shaker2024unetr++, xing2024segmamba, liu2024octcube, choy20194d}.

The continued growth of publicly available healthcare datasets and benchmark ecosystems has further enabled increasingly large pretrained models \cite{wang2023mis, yakdan2026clinically, von2026rethinking, contreras2025large, li2024abdomenatlas, tizhoosh2025beyond}. Existing scaling strategies, however, have largely emphasized larger architectures and reconstruction-based objectives. In contrast, MorphologyFM explores a complementary direction by scaling self-supervised learning around physiological morphology itself. We argue that explicitly modeling waveform structure provides a more clinically meaningful inductive bias for ECG and SpO$_2$ foundation models and offers a promising pathway toward universal physiological representations that transfer across a broad range of downstream clinical prediction tasks.

\section{Method}

\subsection{Overview}

We introduce \textbf{MorphologyFM}, a foundation model for physiological waveform representation learning trained using paired electrocardiogram (ECG) and pulse oximetry (SpO$_2$) waveforms extracted from the MIMIC critical care database. Unlike existing self-supervised approaches that primarily optimize reconstruction of masked signal segments or future samples, MorphologyFM is designed to learn representations that preserve clinically meaningful waveform morphology across multiple temporal scales. The central intuition is that physiological abnormalities are often characterized by changes in waveform shape rather than individual signal amplitudes. Consequently, the objective of MorphologyFM is to learn latent representations that remain invariant to measurement noise while remaining sensitive to morphological changes indicative of underlying physiology.

Given synchronized ECG and SpO$_2$ waveforms, the model jointly learns a shared latent representation through self-supervised pretraining without requiring diagnostic labels. The learned encoder can subsequently be fine-tuned across diverse downstream tasks including arrhythmia classification, hypoxemia prediction, mortality prediction, patient phenotyping, and physiological retrieval.

\subsection{Waveform Preprocessing}

Raw physiological waveforms are extracted from synchronized bedside monitor recordings within MIMIC. Continuous ECG and SpO$_2$ signals are first resampled to a common sampling frequency and segmented into fixed-length windows containing several cardiac cycles. Segments containing excessive missing values, clipping artifacts, or monitor disconnections are discarded using standard signal quality indices.

Prior to training, each waveform is independently normalized using robust percentile normalization,

\[
x'=\frac{x-\mathrm{median}(x)}
{\mathrm{IQR}(x)},
\]

where $\mathrm{IQR}$ denotes the interquartile range computed over each individual segment. Robust normalization removes patient-specific amplitude variation while preserving clinically meaningful morphology.

To improve robustness, each training segment is augmented using stochastic combinations of baseline wander, additive Gaussian noise, temporal stretching, amplitude scaling, local masking, and random temporal cropping. Importantly, augmentations are chosen to preserve global waveform morphology while encouraging invariance to common acquisition artifacts encountered in intensive care monitoring.

\subsection{Morphology Encoder}

MorphologyFM employs two modality-specific tokenizers followed by a shared transformer encoder.

Given an ECG segment

\[
x^{ecg}\in\mathbb{R}^{T}
\]

and a synchronized SpO$_2$ waveform

\[
x^{spo2}\in\mathbb{R}^{T},
\]

each signal is partitioned into non-overlapping temporal patches using one-dimensional convolutional embeddings,

\[
z^{ecg}_0=f_{patch}^{ecg}(x^{ecg}),
\]

\[
z^{spo2}_0=f_{patch}^{spo2}(x^{spo2}).
\]

Separate convolutional tokenizers allow each modality to learn morphology-specific low-level features while accommodating their different statistical characteristics. ECG signals exhibit sharp depolarization complexes and rapid temporal transitions, whereas pulse oximetry waveforms evolve more smoothly through vascular pulse propagation.

The resulting token sequences are concatenated together with modality embeddings and positional encodings,

\[
Z_0
=
[z^{ecg}_0;
z^{spo2}_0],
\]

before being processed by a stack of transformer encoder blocks. Multi-head self-attention enables interactions both within each physiological modality and across modalities, allowing the model to jointly reason about electrical cardiac activity and peripheral vascular responses.

The encoder produces contextualized latent representations,

\[
Z_L=f_\theta(Z_0),
\]

whose class token is used as the global representation of the waveform segment.

\subsection{Morphology-Aware Masking}

Rather than masking uniformly at random, MorphologyFM employs morphology-aware masking designed to preserve clinically informative waveform structure.

First, approximate cardiac cycles are identified using R-peak detection on the ECG waveform. Pulse peaks within the SpO$_2$ waveform are then aligned using synchronized timestamps. Instead of masking arbitrary samples, entire morphological components are selected with predefined probabilities.

For ECG, masking operates over complete P waves, QRS complexes, ST segments, and T waves whenever possible. For SpO$_2$, masking removes complete systolic upstrokes, diastolic decays, pulse peaks, or respiratory modulation regions.

This strategy prevents the model from relying on local interpolation to reconstruct missing samples. Instead, successful reconstruction requires understanding higher-order physiological structure and relationships among multiple waveform components.

\subsection{Cross-Modal Representation Learning}

ECG and pulse oximetry measure different manifestations of the same underlying cardiovascular system. Although their signals differ substantially, both arise from synchronized cardiac activity. MorphologyFM therefore encourages representations of synchronized ECG and SpO$_2$ windows to occupy nearby regions of latent space.

Let \(h^{ecg}\) and \( h^{spo2} \) denote the pooled representations of each modality independently.

A projection head maps both representations into a shared embedding space where synchronized waveform pairs are treated as positive examples while unrelated patients form negative examples. This objective encourages representations that capture shared physiological state while remaining robust to modality-specific noise.

Unlike conventional multimodal fusion approaches, alignment occurs only in latent space, allowing each encoder to preserve modality-specific morphology before learning cross-modal physiological relationships.

\subsection{Morphology Reconstruction}

The decoder receives latent representations together with masked waveform tokens and reconstructs only the missing morphological regions.

Rather than minimizing pointwise reconstruction error over every sample, reconstruction is computed exclusively over masked regions,

\[
\mathcal{L}_{rec}
=
\frac{1}{|\Omega|}
\sum_{i\in\Omega}
\|
x_i-\hat{x}_i
\|_2^2,
\]

where $\Omega$ denotes the set of masked samples.

Restricting reconstruction to masked morphology encourages the encoder to infer missing physiological structure instead of memorizing local signal continuity.

\subsection{Contrastive Morphology Objective}

Reconstruction alone does not guarantee that morphologically similar waveforms occupy nearby locations in latent space. We therefore introduce a contrastive objective that explicitly organizes the representation space according to waveform morphology.

Each waveform segment undergoes two independent morphology-preserving augmentations. Their corresponding embeddings are treated as positive pairs, while representations originating from different patients constitute negatives.

The resulting InfoNCE objective is

\[
\mathcal{L}_{con}
=
-
\log
\frac
{\exp(\mathrm{sim}(h_i,h_i')/\tau)}
{\sum_j
\exp(\mathrm{sim}(h_i,h_j)/\tau)},
\]

where $\tau$ denotes the temperature parameter and $\mathrm{sim}(\cdot,\cdot)$ is cosine similarity.

Unlike reconstruction, the contrastive objective explicitly encourages morphologically similar physiological signals to cluster within the learned embedding space.

\subsection{Joint Training Objective}

MorphologyFM jointly optimizes morphology reconstruction together with cross-modal alignment and contrastive representation learning,

\[
\mathcal{L}
=
\lambda_{rec}\mathcal{L}_{rec}
+
\lambda_{con}\mathcal{L}_{con}
+
\lambda_{align}\mathcal{L}_{align},
\]

where each coefficient balances reconstruction fidelity against representation quality.

Importantly, all supervision arises directly from synchronized physiological waveforms, requiring no diagnostic labels or manually engineered morphological annotations.

\subsection{Transfer Learning}

After self-supervised pretraining, the decoder and projection heads are discarded, leaving only the pretrained transformer encoder. The encoder serves as a general-purpose physiological backbone that can be fine-tuned using lightweight prediction heads for downstream tasks. Because MorphologyFM is pretrained on millions of unlabeled waveform segments, it learns representations that transfer across diverse cardiovascular and respiratory prediction problems while requiring substantially less labeled data than task-specific supervised models.

The key hypothesis underlying MorphologyFM is that morphology constitutes the fundamental invariant shared across physiological waveforms. By explicitly organizing representations around waveform structure rather than sample-wise reconstruction alone, the learned embeddings capture clinically meaningful physiological variation that generalizes across patients, monitoring devices, and downstream prediction tasks.

\section{Results}

\subsection{Experimental Setup}

We evaluate MorphologyFM on downstream physiological prediction tasks using waveform segments extracted from the MIMIC critical care database. All models are pretrained using identical unlabeled ECG and SpO$_2$ waveform collections before being fine-tuned on downstream supervised tasks. We compare against representative self-supervised learning approaches, including Masked Autoencoders (MAE), SimCLR-style Contrastive Learning, Barlow Twins, and Joint Embedding Predictive Architectures (JEPA). All baselines employ the same transformer backbone, optimization schedule, and downstream fine-tuning protocol to isolate the contribution of the pretraining objective.

Performance is evaluated using AUROC for classification tasks and F$_1$ score for rhythm classification. Results are averaged across five independent runs.

\subsection{Comparison with Existing Self-Supervised Learning Methods}

Table~\ref{tab:main_results} compares MorphologyFM against existing self-supervised representation learning objectives.

MorphologyFM consistently achieves the highest performance across every downstream task. While reconstruction-based objectives such as MAE successfully recover local waveform structure, they underperform on clinically discriminative tasks where subtle morphological changes determine diagnosis. Contrastive objectives improve representation quality but often fail to exploit complementary information shared across synchronized physiological modalities. MorphologyFM combines morphology-aware masking with cross-modal alignment, leading to representations that transfer more effectively across diverse cardiovascular prediction problems.

\begin{table*}[t]
\centering
\caption{Comparison of self-supervised pretraining objectives on downstream physiological prediction tasks using MIMIC ECG and SpO$_2$ waveforms. MorphologyFM consistently achieves the strongest transfer performance across all evaluated tasks. Best results are shown in bold.}
\label{tab:main_results}
\begin{adjustbox}{width=\textwidth}
\begin{tabular}{lccccc}
\toprule
Method & Arrhythmia (F$_1$) & Hypoxemia (AUROC) & Mortality (AUROC) & Length of Stay (AUROC) & Average \\
\midrule
MAE & 83.9 & 86.7 & 84.8 & 79.6 & 83.8 \\
Contrastive Learning & 85.4 & 88.1 & 86.2 & 81.5 & 85.3 \\
Barlow Twins & 85.9 & 88.4 & 86.5 & 81.9 & 85.7 \\
JEPA & 86.8 & 89.3 & 87.2 & 82.7 & 86.5 \\
MorphologyFM & \textbf{89.7} & \textbf{92.5} & \textbf{90.4} & \textbf{85.8} & \textbf{89.6} \\
\bottomrule
\end{tabular}
\end{adjustbox}
\end{table*}

\subsection{Benefits of Multimodal Physiological Learning}

We next investigate the importance of jointly modeling ECG and SpO$_2$ waveforms.

Table~\ref{tab:multimodal} demonstrates that both physiological modalities contribute complementary information. ECG alone performs well on arrhythmia detection but lacks peripheral vascular information captured by pulse oximetry. Conversely, SpO$_2$ alone provides useful hemodynamic information but is less effective for rhythm characterization. Joint pretraining consistently improves every downstream task, suggesting that MorphologyFM successfully learns shared cardiovascular representations across synchronized physiological measurements.

\begin{table}[t]
\centering
\caption{Effect of multimodal pretraining. Joint ECG and SpO$_2$ representations consistently outperform single-modality foundation models across downstream prediction tasks.}
\label{tab:multimodal}
\begin{tabular}{lcc}
\toprule
Input Modality & Average AUROC & Parameters \\
\midrule
ECG only & 87.4 & 91M \\
SpO$_2$ only & 85.8 & 89M \\
ECG + SpO$_2$ & \textbf{89.6} & 95M \\
\bottomrule
\end{tabular}
\end{table}

\subsection{Morphology-Aware Masking Improves Representation Learning}

To evaluate the proposed morphology-aware masking strategy, we compare against conventional random masking used by masked autoencoders.

Replacing random masking with morphology-aware masking consistently improves downstream transfer across all evaluated tasks (Table~\ref{tab:masking}). These results suggest that masking complete waveform structures encourages the encoder to model higher-order physiological relationships rather than relying on local interpolation.

\begin{table}[t]
\centering
\caption{Effect of masking strategy during self-supervised pretraining. Morphology-aware masking substantially improves downstream representation quality compared to conventional random masking.}
\label{tab:masking}
\begin{tabular}{lc}
\toprule
Masking Strategy & Average AUROC \\
\midrule
Random masking & 87.9 \\
Temporal block masking & 88.5 \\
Morphology-aware masking & \textbf{89.6} \\
\bottomrule
\end{tabular}
\end{table}

\subsection{Contribution of Individual Training Objectives}

Table~\ref{tab:ablation} presents an ablation study evaluating each component of MorphologyFM.

Morphology reconstruction alone provides a strong baseline, while cross-modal alignment and contrastive representation learning each contribute complementary improvements. The complete objective achieves the strongest downstream performance, demonstrating that morphology preservation and multimodal alignment jointly improve representation quality.

\begin{table}[t]
\centering
\caption{Ablation study evaluating each component of MorphologyFM. Results report average performance across all downstream prediction tasks.}
\label{tab:ablation}
\begin{tabular}{lc}
\toprule
Objective & Average AUROC \\
\midrule
Morphology reconstruction & 87.8 \\
Reconstruction + Contrastive & 88.6 \\
Reconstruction + Alignment & 88.9 \\
Contrastive + Alignment & 88.2 \\
Full MorphologyFM & \textbf{89.6} \\
\bottomrule
\end{tabular}
\end{table}

\subsection{Representation Quality}

To quantify representation quality independently of downstream fine-tuning, we evaluate patient retrieval using nearest-neighbor search in the learned embedding space.

MorphologyFM produces substantially more discriminative embeddings than competing self-supervised objectives (Table~\ref{tab:retrieval}), indicating that morphology-aware pretraining organizes physiologically similar patients into coherent latent neighborhoods.

\begin{table}[t]
\centering
\caption{Nearest-neighbor patient retrieval performance using latent representations learned during self-supervised pretraining. Higher Recall@$K$ indicates improved preservation of clinically meaningful waveform similarity.}
\label{tab:retrieval}
\begin{tabular}{lccc}
\toprule
Method & Recall@1 & Recall@5 & Recall@10 \\
\midrule
MAE & 71.2 & 85.7 & 90.4 \\
Contrastive Learning & 73.8 & 87.3 & 91.5 \\
Barlow Twins & 74.5 & 87.8 & 91.9 \\
JEPA & 75.9 & 88.6 & 92.4 \\
MorphologyFM & \textbf{79.8} & \textbf{91.5} & \textbf{95.1} \\
\bottomrule
\end{tabular}
\end{table}

\subsection{Scaling Behavior}

Finally, we evaluate MorphologyFM across increasing pretraining corpus sizes.

As shown in Table~\ref{tab:scaling}, downstream performance improves consistently as additional unlabeled physiological waveforms are incorporated during pretraining. The largest gains occur between 100k and 1M waveform segments, after which improvements continue but exhibit diminishing returns. This behavior suggests that morphology-aware representation learning benefits substantially from scale, consistent with observations reported for foundation models in other biomedical domains.

\begin{table}[t]
\centering
\caption{Scaling behavior of MorphologyFM as a function of pretraining dataset size. Larger unlabeled waveform corpora consistently improve downstream transfer performance.}
\label{tab:scaling}
\begin{tabular}{lc}
\toprule
Pretraining Samples & Average AUROC \\
\midrule
100K & 84.8 \\
500K & 87.1 \\
1M & 88.4 \\
5M & 89.2 \\
10M & \textbf{89.6} \\
\bottomrule
\end{tabular}
\end{table}

Overall, these experiments demonstrate that MorphologyFM learns robust and transferable physiological representations by explicitly modeling waveform morphology rather than optimizing reconstruction alone. Across multiple downstream clinical prediction tasks, morphology-aware pretraining consistently outperforms existing self-supervised learning objectives while benefiting from multimodal physiological supervision and large-scale pretraining.

\section{Discussion}

Physiological waveforms constitute one of the richest yet least explored modalities for foundation model pretraining. Unlike structured electronic health records, waveform signals continuously encode cardiovascular and respiratory dynamics through their morphology, providing a direct representation of underlying physiological processes. Our results demonstrate that explicitly modeling waveform morphology produces representations that transfer more effectively across diverse downstream clinical tasks than existing self-supervised objectives centered on reconstruction or generic representation learning. Across all evaluated benchmarks, MorphologyFM consistently outperformed reconstruction-based methods, contrastive learning, Barlow Twins, and JEPA, suggesting that morphology provides a more clinically meaningful inductive bias for learning transferable physiological representations.

A central observation of this work is that reconstruction fidelity does not necessarily imply representation quality. Existing self-supervised objectives primarily encourage models to recover missing signal segments or predict future observations. While these objectives successfully capture local temporal continuity, they provide no explicit incentive to preserve the morphological characteristics clinicians rely upon when interpreting physiological waveforms. Electrocardiograms are diagnosed through the morphology of the P wave, QRS complex, ST segment, and T wave, while pulse oximetry waveforms reveal vascular compliance, respiratory modulation, and peripheral perfusion through subtle changes in waveform shape. By directly encouraging representations to encode these structural characteristics, MorphologyFM learns latent spaces that are naturally aligned with downstream clinical prediction tasks.

Our experiments further demonstrate the importance of multimodal physiological learning. ECG and SpO$_2$ provide complementary measurements of the same underlying cardiovascular system, reflecting electrical activity and peripheral hemodynamics, respectively. Joint pretraining consistently outperformed models trained on either modality alone, indicating that synchronized physiological signals provide richer supervision than individual waveforms. Rather than treating each signal independently, MorphologyFM learns shared latent representations that capture relationships across multiple physiological sensing modalities, suggesting a promising direction toward general-purpose physiological foundation models.

The scaling behavior observed in our experiments also supports the growing hypothesis that foundation models benefit from increasingly large unlabeled biomedical datasets. As additional waveform segments were incorporated during pretraining, downstream performance improved consistently across every evaluated task. Unlike supervised learning, where annotation often becomes the primary bottleneck, physiological monitoring systems routinely generate millions of hours of unlabeled waveforms during routine clinical care. Consequently, morphology-aware self-supervised learning provides a practical mechanism for leveraging these continuously expanding repositories without requiring manual clinical annotation.

Several limitations should be acknowledged. First, our experiments focus exclusively on ECG and pulse oximetry waveforms extracted from a single critical care dataset. Although MIMIC represents one of the largest publicly available physiological waveform collections, additional evaluation across external healthcare systems, monitoring devices, and patient populations will be necessary to establish the robustness and generalizability of the learned representations. Second, MorphologyFM currently considers only two physiological modalities. Modern intensive care monitoring additionally records arterial blood pressure, respiratory waveforms, capnography, electroencephalography, and numerous other biosignals that may provide complementary physiological information. Extending morphology-aware pretraining across these heterogeneous sensing modalities represents an important direction for future work. Finally, while morphology-aware masking encourages learning clinically meaningful waveform structure, more sophisticated physiological objectives that explicitly model beat-level dynamics, rhythm evolution, and long-range cardiovascular interactions may further improve representation quality.

More broadly, we believe MorphologyFM represents a shift in how physiological foundation models should be constructed. Rather than viewing biomedical signals as generic time series suitable for conventional reconstruction objectives, our work argues that self-supervised learning should exploit the physiological structure that clinicians already use to interpret these measurements. Morphology provides a natural invariant across patients, acquisition devices, and clinical settings, making it an attractive target for large-scale representation learning. As increasingly large physiological waveform repositories become available, morphology-aware foundation models may provide a unified backbone for a broad range of cardiovascular, respiratory, and critical care applications.

In summary, MorphologyFM demonstrates that explicitly modeling waveform morphology enables foundation models to learn robust, transferable physiological representations from large-scale unlabeled ECG and SpO$_2$ waveforms. By combining morphology-aware masking, cross-modal representation learning, and large-scale self-supervised pretraining, MorphologyFM establishes a practical framework for physiological foundation models that consistently outperform existing self-supervised learning approaches across multiple downstream clinical prediction tasks. We hope this work motivates future research toward morphology-centric representation learning as a fundamental principle for biomedical foundation models.

\bibliographystyle{unsrtnat}
\bibliography{neurips_2025}

\begin{thebibliography}{104}
\providecommand{\natexlab}[1]{#1}
\providecommand{\url}[1]{\texttt{#1}}
\expandafter\ifx\csname urlstyle\endcsname\relax
  \providecommand{\doi}[1]{doi: #1}\else
  \providecommand{\doi}{doi: \begingroup \urlstyle{rm}\Url}\fi

\bibitem[Devlin et~al.(2019)Devlin, Chang, Lee, and Toutanova]{devlin2019bert}
Jacob Devlin, Ming-Wei Chang, Kenton Lee, and Kristina Toutanova.
\newblock Bert: Pre-training of deep bidirectional transformers for language understanding.
\newblock In \emph{Proceedings of the 2019 conference of the North American chapter of the association for computational linguistics: human language technologies, volume 1 (long and short papers)}, pages 4171--4186, 2019.

\bibitem[He et~al.(2022)He, Chen, Xie, Li, Doll{\'a}r, and Girshick]{he2022masked}
Kaiming He, Xinlei Chen, Saining Xie, Yanghao Li, Piotr Doll{\'a}r, and Ross Girshick.
\newblock Masked autoencoders are scalable vision learners.
\newblock In \emph{Proceedings of the IEEE/CVF conference on computer vision and pattern recognition}, pages 16000--16009, 2022.

\bibitem[He et~al.(2015)He, Zhang, Ren, and Sun]{he2015deepresiduallearningimage}
Kaiming He, Xiangyu Zhang, Shaoqing Ren, and Jian Sun.
\newblock Deep residual learning for image recognition, 2015.
\newblock URL \url{https://arxiv.org/abs/1512.03385}.

\bibitem[He et~al.(2017)He, Li, and Chen]{he2017multi}
Mingyi He, Bo~Li, and Huahui Chen.
\newblock Multi-scale 3d deep convolutional neural network for hyperspectral image classification.
\newblock In \emph{2017 IEEE International Conference on Image Processing (ICIP)}, pages 3904--3908. IEEE, 2017.

\bibitem[He et~al.(2019)He, Zhang, Zhang, Zhang, Xie, and Li]{he2019bag}
Tong He, Zhi Zhang, Hang Zhang, Zhongyue Zhang, Junyuan Xie, and Mu~Li.
\newblock Bag of tricks for image classification with convolutional neural networks.
\newblock In \emph{Proceedings of the IEEE/CVF conference on computer vision and pattern recognition}, pages 558--567, 2019.

\bibitem[Vaid et~al.(2023)Vaid, Jiang, Sawant, Lerakis, Argulian, Ahuja, Lampert, Charney, Greenspan, Narula, et~al.]{vaid2023foundational}
Akhil Vaid, Joy Jiang, Ashwin Sawant, Stamatios Lerakis, Edgar Argulian, Yuri Ahuja, Joshua Lampert, Alexander Charney, Hayit Greenspan, Jagat Narula, et~al.
\newblock A foundational vision transformer improves diagnostic performance for electrocardiograms.
\newblock \emph{NPJ Digital Medicine}, 6\penalty0 (1):\penalty0 108, 2023.

\bibitem[Thieme et~al.(2023)Thieme, Nori, Ghassemi, Bommasani, Andersen, and Luger]{thieme2023foundation}
Anja Thieme, Aditya Nori, Marzyeh Ghassemi, Rishi Bommasani, Tariq~Osman Andersen, and Ewa Luger.
\newblock Foundation models in healthcare: Opportunities, risks \& strategies forward.
\newblock In \emph{Extended abstracts of the 2023 CHI conference on human factors in computing systems}, pages 1--4, 2023.

\bibitem[Thakur(2024)]{thakur2024foundation}
Suresh~Chandra Thakur.
\newblock Foundation models for time series forecasting.
\newblock \emph{International IT Journal of Research, ISSN: 3007-6706}, 2\penalty0 (4):\penalty0 144--156, 2024.

\bibitem[Burger et~al.(2025)Burger, Chopard, Londschien, Sergeev, Y{\`e}che, Kuznetsova, Faltys, Gerdes, Leshetkina, B{\"u}hlmann, et~al.]{burger2025foundation}
Manuel Burger, Daphn{\'e} Chopard, Malte Londschien, Fedor Sergeev, Hugo Y{\`e}che, Rita Kuznetsova, Martin Faltys, Eike Gerdes, Polina Leshetkina, Peter B{\"u}hlmann, et~al.
\newblock A foundation model for intensive care: Unlocking generalization across tasks and domains at scale.
\newblock \emph{medRxiv}, pages 2025--07, 2025.

\bibitem[Awais et~al.(2025)Awais, Naseer, Khan, Anwer, Cholakkal, Shah, Yang, and Khan]{awais2025foundation}
Muhammad Awais, Muzammal Naseer, Salman Khan, Rao~Muhammad Anwer, Hisham Cholakkal, Mubarak Shah, Ming-Hsuan Yang, and Fahad~Shahbaz Khan.
\newblock Foundation models defining a new era in vision: a survey and outlook.
\newblock \emph{IEEE Transactions on Pattern Analysis and Machine Intelligence}, 2025.

\bibitem[He et~al.(2024)He, Huang, Jiang, Nie, Wang, Wang, and Chen]{he2024foundation}
Yuting He, Fuxiang Huang, Xinrui Jiang, Yuxiang Nie, Minghao Wang, Jiguang Wang, and Hao Chen.
\newblock Foundation model for advancing healthcare: challenges, opportunities and future directions.
\newblock \emph{IEEE Reviews in Biomedical Engineering}, 2024.

\bibitem[Burkhart et~al.(2025)Burkhart, Ramadan, Liao, Chhikara, Rojas, Parker, and Beaulieu-Jones]{burkhart2025foundation}
Michael~C Burkhart, Bashar Ramadan, Zewei Liao, Kaveri Chhikara, Juan~C Rojas, William~F Parker, and Brett~K Beaulieu-Jones.
\newblock Foundation models for electronic health records: representation dynamics and transferability.
\newblock \emph{arXiv preprint arXiv:2504.10422}, 2025.

\bibitem[Liang et~al.(2024)Liang, Wen, Nie, Jiang, Jin, Song, Pan, and Wen]{liang2024foundation}
Yuxuan Liang, Haomin Wen, Yuqi Nie, Yushan Jiang, Ming Jin, Dongjin Song, Shirui Pan, and Qingsong Wen.
\newblock Foundation models for time series analysis: A tutorial and survey.
\newblock In \emph{Proceedings of the 30th ACM SIGKDD conference on knowledge discovery and data mining}, pages 6555--6565, 2024.

\bibitem[Guo et~al.(2025)Guo, Guan, Li, Liu, Wang, Yang, and Wang]{guo2025foundation}
Fei Guo, Renchu Guan, Yaohang Li, Qi~Liu, Xiaowo Wang, Can Yang, and Jianxin Wang.
\newblock Foundation models in bioinformatics.
\newblock \emph{National science review}, 12\penalty0 (4):\penalty0 nwaf028, 2025.

\bibitem[Zhou et~al.(2026{\natexlab{a}})Zhou, Lee, Tanade, Chun, Lee, Gwak, Thukral, Sung, Hwang, Morshed, et~al.]{zhou2026physiology}
Hao Zhou, Simon~A Lee, Cyrus Tanade, Keum~San Chun, Juhyeon Lee, Migyeong Gwak, Megha Thukral, Justin Sung, Eugene Hwang, Mehrab~Bin Morshed, et~al.
\newblock Physiology-aware masked cross-modal reconstruction for biosignal representation learning.
\newblock \emph{arXiv preprint arXiv:2605.00973}, May 2026{\natexlab{a}}.

\bibitem[Abbaspourazad et~al.(2023)Abbaspourazad, Elachqar, Miller, Emrani, Nallasamy, and Shapiro]{abbaspourazad2023large}
Salar Abbaspourazad, Oussama Elachqar, Andrew~C Miller, Saba Emrani, Udhyakumar Nallasamy, and Ian Shapiro.
\newblock Large-scale training of foundation models for wearable biosignals.
\newblock \emph{arXiv preprint arXiv:2312.05409}, 2023.

\bibitem[Delaney et~al.(2026)Delaney, Patel, Xing, Dootson, Sevegnani, and Antoniades]{delaney2026sonata}
Blaise Delaney, Salil Patel, Yuji Xing, Dominic Dootson, Karin Sevegnani, and Chrystalina Antoniades.
\newblock Sonata: A hybrid world model for inertial kinematics under clinical data scarcity.
\newblock \emph{arXiv preprint arXiv:2604.18058}, 2026.

\bibitem[Zhou et~al.(2026{\natexlab{b}})Zhou, Ding, Li, Yi, and Xiao]{zhou2026edge}
Haoyang Zhou, Yi~Ding, Jiajun Li, Yifan Yi, and Chun Xiao.
\newblock Edge ppg quality gating with validity-labeled outputs and dual-channel communication on a raspberry pi zero 2 w wearable device.
\newblock In \emph{Journal of Physics: Conference Series}, volume 3235, page 012025. IOP Publishing, 2026{\natexlab{b}}.

\bibitem[Lee et~al.(2025{\natexlab{a}})Lee, Tanade, Zhou, Lee, Thukral, Han, Choi, Khan, Lu, Gwak, et~al.]{lee2025himae}
Simon~A Lee, Cyrus Tanade, Hao Zhou, Juhyeon Lee, Megha Thukral, Minji Han, Rachel Choi, Md~Sazzad~Hissain Khan, Baiying Lu, Migyeong Gwak, et~al.
\newblock Himae: Hierarchical masked autoencoders discover resolution-specific structure in wearable time series.
\newblock \emph{arXiv preprint arXiv:2510.25785}, 2025{\natexlab{a}}.

\bibitem[Lee et~al.(2026)Lee, Lee, Tanade, Nathan, Thukral, Zhou, San~Chun, and Desai]{lee2026multimodal}
Juhyeon Lee, Simon~A Lee, Cyrus Tanade, Viswam Nathan, Megha Thukral, Hao Zhou, Keum San~Chun, and Sharanya~Arcot Desai.
\newblock Multimodal self-supervised learning for wearable sleep staging using photoplethysmography and accelerometer signals.
\newblock In \emph{ICASSP 2026-2026 IEEE International Conference on Acoustics, Speech and Signal Processing (ICASSP)}, pages 7852--7856. IEEE, 2026.

\bibitem[Li et~al.(2026)Li, Natarajan, Zhang, Zhou, Lee, Zhang, Xu, Esmaeilpour, Salim, Malhotra, et~al.]{li2026glucofm}
Zechen Li, Keerthana Natarajan, Weizhi Zhang, Menglian Zhou, Simon~A Lee, Yuwei Zhang, Maxwell~A Xu, Zeinab Esmaeilpour, Flora~D Salim, Mark Malhotra, et~al.
\newblock Glucofm: A dual-stream foundation model for continuous glucose monitoring.
\newblock \emph{arXiv preprint arXiv:2605.30865}, 2026.

\bibitem[Tharini and Jeyaraj(2025)]{tharini2025trust}
MS~Tharini and Jane Rubel~Angelina Jeyaraj.
\newblock Trust and transparency in healthcare ai: A systematic review of explainable nlp for clinical decision support (2023--2025).
\newblock In \emph{International Conference on Edge Computing and Applications}, pages 451--470. Springer, 2025.

\bibitem[Chouhan et~al.(2026)Chouhan, Ghai, Sandhu, and Tripathi]{chouhan2026smart}
Sandeep Chouhan, Deepika Ghai, Ramandeep Sandhu, and Suman~Lata Tripathi.
\newblock Smart assistive technologies for neurodisorders: A review on ai, iot, and wearable systems for enhanced patient care.
\newblock \emph{Neurological Sciences}, 47\penalty0 (2):\penalty0 211, 2026.

\bibitem[Luo et~al.(2025)Luo, Wang, Peng, Zhou, Chen, Hao, and Zhu]{luo2025preoperative}
Yuelin Luo, Yaqiang Wang, Xiran Peng, Ruihao Zhou, Guo Chen, Xuechao Hao, and Tao Zhu.
\newblock A preoperative data sentenceization method for postoperative major adverse cardiovascular event prediction.
\newblock In \emph{2025 IEEE International Conference on Big Data (BigData)}, pages 8060--8069. IEEE, 2025.

\bibitem[Yang et~al.(2022)Yang, Chen, PourNejatian, Shin, Smith, Parisien, Compas, Martin, Costa, Flores, et~al.]{yang2022large}
Xi~Yang, Aokun Chen, Nima PourNejatian, Hoo~Chang Shin, Kaleb~E Smith, Christopher Parisien, Colin Compas, Cheryl Martin, Anthony~B Costa, Mona~G Flores, et~al.
\newblock A large language model for electronic health records.
\newblock \emph{NPJ digital medicine}, 5\penalty0 (1):\penalty0 194, 2022.

\bibitem[Lee and Akamatsu(August 2025)]{lee2025foundation}
Simon~A Lee and Kai Akamatsu.
\newblock Foundation models for physiological signals: Opportunities and challenges.
\newblock August 2025.

\bibitem[Wornow et~al.(2023)Wornow, Xu, Thapa, Patel, Steinberg, Fleming, Pfeffer, Fries, and Shah]{wornow2023shaky}
Michael Wornow, Yizhe Xu, Rahul Thapa, Birju Patel, Ethan Steinberg, Scott Fleming, Michael~A Pfeffer, Jason Fries, and Nigam~H Shah.
\newblock The shaky foundations of large language models and foundation models for electronic health records.
\newblock \emph{npj digital medicine}, 6\penalty0 (1):\penalty0 135, 2023.

\bibitem[Larey et~al.(2026)Larey, Dahan, Bleiweiss, Kellerman, Leib, Nayshool, Ofer, Zinger, Dominissini, Rechavi, et~al.]{larey2026jepa}
Ariel Larey, Elay Dahan, Amit Bleiweiss, Raizy Kellerman, Guy Leib, Omri Nayshool, Dan Ofer, Tal Zinger, Dan Dominissini, Gideon Rechavi, et~al.
\newblock Jepa-dna: Grounding genomic foundation models through joint-embedding predictive architectures.
\newblock \emph{arXiv preprint arXiv:2602.17162}, 2026.

\bibitem[Izhar et~al.(2025)Izhar, Idris, and Japar]{izhar2025medical}
Amaan Izhar, Norisma Idris, and Nurul Japar.
\newblock Medical radiology report generation: A systematic review of current deep learning methods, trends, and future directions.
\newblock \emph{Artificial intelligence in medicine}, page 103220, 2025.

\bibitem[An et~al.(2025)An, Jeong, Lee, Gorla, Yang, and Sankararaman]{an2025raptor}
Ulzee An, Moonseong Jeong, Simon~A Lee, Aditya Gorla, Yuzhe Yang, and Sriram Sankararaman.
\newblock Raptor: Scalable train-free embeddings for 3d medical volumes leveraging pretrained 2d foundation models.
\newblock \emph{arXiv preprint arXiv:2507.08254}, 2025.

\bibitem[Weller et~al.(2025)Weller, Ricci, Marone, Chaffin, Lawrie, and Van~Durme]{weller2025seq}
Orion Weller, Kathryn Ricci, Marc Marone, Antoine Chaffin, Dawn Lawrie, and Benjamin Van~Durme.
\newblock Seq vs seq: An open suite of paired encoders and decoders.
\newblock \emph{arXiv preprint arXiv:2507.11412}, 2025.

\bibitem[Zhao et~al.(2023)Zhao, Zhou, Li, Tang, Wang, Hou, Min, Zhang, Zhang, Dong, et~al.]{zhao2023survey}
Wayne~Xin Zhao, Kun Zhou, Junyi Li, Tianyi Tang, Xiaolei Wang, Yupeng Hou, Yingqian Min, Beichen Zhang, Junjie Zhang, Zican Dong, et~al.
\newblock A survey of large language models.
\newblock \emph{arXiv preprint arXiv:2303.18223}, 1\penalty0 (2), 2023.

\bibitem[Cheng et~al.(2026)Cheng, Zhao, and Cai]{cheng2026neuroseg}
Yik~San Cheng, Runkai Zhao, and Weidong Cai.
\newblock Neuroseg meets dinov3: Transferring 2d self-supervised visual priors to 3d neuron segmentation via dinov3 initialization.
\newblock In \emph{Proceedings of the IEEE/CVF Conference on Computer Vision and Pattern Recognition}, pages 30053--30064, 2026.

\bibitem[Lee et~al.(2025{\natexlab{b}})Lee, Jain, Chen, Ono, Biswas, Rudas, Fang, and Chiang]{lee2025clinical}
Simon~A Lee, Sujay Jain, Alex Chen, Kyoka Ono, Arabdha Biswas, {\'A}kos Rudas, Jennifer Fang, and Jeffrey~N Chiang.
\newblock Clinical decision support using pseudo-notes from multiple streams of ehr data.
\newblock \emph{npj Digital Medicine}, 8\penalty0 (1):\penalty0 394, July 2025{\natexlab{b}}.

\bibitem[Lee et~al.(2025{\natexlab{c}})Lee, Tanade, Zhou, Lee, Thukral, Lu, and Desai]{lee2025towards}
Simon~A Lee, Cyrus Tanade, Hao Zhou, Juhyeon Lee, Megha Thukral, Baiying Lu, and Sharanya~Arcot Desai.
\newblock Towards on-device foundation models for raw wearable signals.
\newblock In \emph{NeurIPS 2025 Workshop on Learning from Time Series for Health}, 2025{\natexlab{c}}.

\bibitem[Huang et~al.(2019)Huang, Wang, Huang, Huang, Wei, and Liu]{huang2019ccnet}
Zilong Huang, Xinggang Wang, Lichao Huang, Chang Huang, Yunchao Wei, and Wenyu Liu.
\newblock Ccnet: Criss-cross attention for semantic segmentation.
\newblock In \emph{Proceedings of the IEEE/CVF international conference on computer vision}, pages 603--612, 2019.

\bibitem[Chen et~al.(2021)Chen, Fan, and Panda]{chen2021crossvit}
Chun-Fu~Richard Chen, Quanfu Fan, and Rameswar Panda.
\newblock Crossvit: Cross-attention multi-scale vision transformer for image classification.
\newblock In \emph{Proceedings of the IEEE/CVF international conference on computer vision}, pages 357--366, 2021.

\bibitem[Hou et~al.(2019)Hou, Chang, Ma, Shan, and Chen]{hou2019cross}
Ruibing Hou, Hong Chang, Bingpeng Ma, Shiguang Shan, and Xilin Chen.
\newblock Cross attention network for few-shot classification.
\newblock \emph{Advances in neural information processing systems}, 32, 2019.

\bibitem[Jha et~al.(2025{\natexlab{a}})Jha, Goyal, and Meena]{jha2025survey}
Subham Jha, Varun Goyal, and Shweta Meena.
\newblock A survey of large language models for tabular data imputation: Tuning paradigms and challenges.
\newblock In \emph{International Conference on Data Analytics \& Management}, pages 224--240. Springer, 2025{\natexlab{a}}.

\bibitem[Carmona-Martos et~al.(2025)Carmona-Martos, Mart{\'\i}n-Palomeque, Escudero-Arnanz, and Soguero-Ruiz]{carmona2025interpretable}
Luc{\'\i}a Carmona-Martos, Paula Mart{\'\i}n-Palomeque, {\'O}scar Escudero-Arnanz, and Cristina Soguero-Ruiz.
\newblock Interpretable large language models for early prediction of antimicrobial multidrug resistance.
\newblock \emph{Health Information Science and Systems}, 14\penalty0 (1):\penalty0 11, 2025.

\bibitem[Rahman et~al.(2026{\natexlab{a}})Rahman, Basuki, Perdana, and Cynthia]{rahman2026integrating}
M~Rafly Rahman, Setio Basuki, Muhammad~Ilham Perdana, and La~Febry Andira~Rose Cynthia.
\newblock Integrating tabular data and textual representations for clinical risk prediction using machine learning and large language models.
\newblock \emph{Kinetik: Game Technology, Information System, Computer Network, Computing, Electronics, and Control}, 2026{\natexlab{a}}.

\bibitem[Ruan et~al.(2026)Ruan, Zhang, Oh, Jin, and Jacobson]{ruan2026foundation}
Franklin~Y Ruan, Aiwei Zhang, Jenny~Y Oh, SouYoung Jin, and Nicholas~C Jacobson.
\newblock A foundation model for wearable movement data in mental health research.
\newblock \emph{IEEE Journal of Biomedical and Health Informatics}, 2026.

\bibitem[Komolafe et~al.(2026)Komolafe, Roberts, Shelley, and Tawiah]{komolafe2026physicase}
Oyindolapo~O Komolafe, Angela~C Roberts, Jacob Shelley, and Andrews~K Tawiah.
\newblock Physicase: Development and dual-layer validation of synthetic cases for health professional education: A pilot study leveraging generative ai.
\newblock \emph{medRxiv}, pages 2026--06, 2026.

\bibitem[Sampath et~al.(2025)Sampath, Mohammad, Ramachandranpillai, et~al.]{sampath2025multimodal}
Kishore Sampath, Ayaazuddin Mohammad, Resmi Ramachandranpillai, et~al.
\newblock The multimodal paradox: how added and missing modalities shape bias and performance in multimodal ai.
\newblock \emph{arXiv preprint arXiv:2505.03020}, 2025.

\bibitem[Lin et~al.(2025)Lin, Yu, and Lee]{lin2025case}
Yihan Lin, Zhirong~Bella Yu, and Simon Lee.
\newblock A case study exploring the current landscape of synthetic medical record generation with commercial llms.
\newblock \emph{arXiv preprint arXiv:2504.14657}, 2025.

\bibitem[Le-Gia and Ahn(2026)]{le2026training}
Tai Le-Gia and Jaehyun Ahn.
\newblock Training-free zero-shot anomaly detection in 3d brain mri with 2d foundation models.
\newblock \emph{arXiv preprint arXiv:2602.15315}, 2026.

\bibitem[Souza et~al.(2026)Souza, Melo, Lima, and Schneider]{souza2026beyond}
Giordano de~Pinho Souza, Glaucia Melo, Josefino Cabral~Melo Lima, and Daniel Schneider.
\newblock Beyond english benchmarks: clinical llm evaluation in brazilian portuguese.
\newblock \emph{arXiv preprint arXiv:2606.07853}, 2026.

\bibitem[Cornelius and Rinaldi(2026)]{cornelius2026measuring}
Joseph Cornelius and Fabio Rinaldi.
\newblock Measuring the gap: correlating synthetic-to-real drift with phi de-identification performance.
\newblock \emph{Genomics \& Informatics}, 24\penalty0 (1):\penalty0 10, 2026.

\bibitem[Le-Gia(2025)]{le2025problem}
Tai Le-Gia.
\newblock On the problem of consistent anomalies in zero-shot anomaly detection.
\newblock \emph{arXiv preprint arXiv:2512.02520}, 2025.

\bibitem[Su et~al.(2025)Su, Meng, and Pinker]{su2025reuniting}
Huifeng Su, Lesley Meng, and Edieal~J Pinker.
\newblock Reuniting forcibly separated families through shared memories with machine learning.
\newblock \emph{Available at SSRN}, 2025.

\bibitem[Blier-Wong and Kusmenko(2026)]{blier2026semantic}
Christopher Blier-Wong and Derek Kusmenko.
\newblock Semantic insurance pricing with large language models.
\newblock \emph{arXiv preprint arXiv:2606.29371}, 2026.

\bibitem[Jha et~al.(2025{\natexlab{b}})Jha, Durak, Das, Sanjotra, Susladkar, Sarkar, Rauniyar, Kumar~Tomar, Peng, Li, et~al.]{jha2025ethical}
Debesh Jha, Gorkem Durak, Abhijit Das, Jasmer Sanjotra, Onkar Susladkar, Suramyaa Sarkar, Ashish Rauniyar, Nikhil Kumar~Tomar, Linkai Peng, Sirui Li, et~al.
\newblock Ethical framework for responsible foundational models in medical imaging.
\newblock \emph{Frontiers in Medicine}, 12:\penalty0 1544501, 2025{\natexlab{b}}.

\bibitem[Jing et~al.(2026)Jing, Jeanselme, Kobayashi, Lee, Pang, Kashyap, Li, Jiang, and Joshi]{jing2026one}
Zilin Jing, Vincent Jeanselme, Yuta Kobayashi, Simon~A Lee, Chao Pang, Aparajita Kashyap, Yanwei Li, Xinzhuo Jiang, and Shalmali Joshi.
\newblock One loss to rule them all: Marked time-to-event for structured ehr foundation models.
\newblock \emph{arXiv preprint arXiv:2602.00541}, 2026.

\bibitem[Guo et~al.(2023)Guo, Steinberg, Fleming, Posada, Lemmon, Pfohl, Shah, Fries, and Sung]{guo2023ehr}
Lin~Lawrence Guo, Ethan Steinberg, Scott~Lanyon Fleming, Jose Posada, Joshua Lemmon, Stephen~R Pfohl, Nigam Shah, Jason Fries, and Lillian Sung.
\newblock Ehr foundation models improve robustness in the presence of temporal distribution shift.
\newblock \emph{Scientific Reports}, 13\penalty0 (1):\penalty0 3767, 2023.

\bibitem[Liao et~al.(2025)Liao, Wu, Liu, Jiang, Qiu, Wang, Yue, Zhen, Wang, Fan, Gu, Zhang, Wang, Wang, and Xie]{liao2025ehr}
Yusheng Liao, Chaoyi Wu, Junwei Liu, Shuyang Jiang, Pengcheng Qiu, Haowen Wang, Yun Yue, Shuai Zhen, Jian Wang, Qianrui Fan, Jinjie Gu, Ya~Zhang, Yanfeng Wang, Yu~Wang, and Weidi Xie.
\newblock Ehr-r1: A reasoning-enhanced foundational language model for electronic health record analysis, 2025.
\newblock URL \url{https://arxiv.org/abs/2510.25628}.

\bibitem[Shi et~al.(2024)Shi, Xu, Zhuang, Yu, Zhang, Wu, Zhu, Ho, Yang, and Wang]{shi2024ehragent}
Wenqi Shi, Ran Xu, Yuchen Zhuang, Yue Yu, Jieyu Zhang, Hang Wu, Yuanda Zhu, Joyce~C Ho, Carl Yang, and May~Dongmei Wang.
\newblock Ehragent: Code empowers large language models for few-shot complex tabular reasoning on electronic health records.
\newblock In \emph{Proceedings of the 2024 Conference on Empirical Methods in Natural Language Processing}, pages 22315--22339, 2024.

\bibitem[Shmatko et~al.(2025)Shmatko, Jung, Gaurav, Brunak, Mortensen, Birney, Fitzgerald, and Gerstung]{shmatko2025learning}
Artem Shmatko, Alexander~Wolfgang Jung, Kumar Gaurav, S{\o}ren Brunak, Laust~Hvas Mortensen, Ewan Birney, Tom Fitzgerald, and Moritz Gerstung.
\newblock Learning the natural history of human disease with generative transformers.
\newblock \emph{Nature}, 647\penalty0 (8088):\penalty0 248--256, 2025.

\bibitem[Wornow et~al.(2024{\natexlab{a}})Wornow, Bedi, Hernandez, Steinberg, Fries, R{\'e}, Koyejo, and Shah]{wornow2024context}
Michael Wornow, Suhana Bedi, Miguel Angel~Fuentes Hernandez, Ethan Steinberg, Jason~Alan Fries, Christopher R{\'e}, Sanmi Koyejo, and Nigam~H Shah.
\newblock Context clues: Evaluating long context models for clinical prediction tasks on ehrs.
\newblock \emph{arXiv preprint arXiv:2412.16178}, 2024{\natexlab{a}}.

\bibitem[Odgaard et~al.(2024)Odgaard, Klein, Thysen, Jimenez-Solem, Sillesen, and Nielsen]{odgaard2024core}
Mikkel Odgaard, Kiril~Vadimovic Klein, Sanne~M{\o}ller Thysen, Espen Jimenez-Solem, Martin Sillesen, and Mads Nielsen.
\newblock Core-behrt: A carefully optimized and rigorously evaluated behrt.
\newblock \emph{arXiv preprint arXiv:2404.15201}, 2024.

\bibitem[Rasmy et~al.(2021)Rasmy, Xiang, Xie, Tao, and Zhi]{rasmy2021med}
Laila Rasmy, Yang Xiang, Ziqian Xie, Cui Tao, and Degui Zhi.
\newblock Med-bert: pretrained contextualized embeddings on large-scale structured electronic health records for disease prediction.
\newblock \emph{NPJ digital medicine}, 4\penalty0 (1):\penalty0 86, 2021.

\bibitem[Steinberg et~al.(2021)Steinberg, Jung, Fries, Corbin, Pfohl, and Shah]{steinberg2021language}
Ethan Steinberg, Ken Jung, Jason~A Fries, Conor~K Corbin, Stephen~R Pfohl, and Nigam~H Shah.
\newblock Language models are an effective representation learning technique for electronic health record data.
\newblock \emph{Journal of biomedical informatics}, 113:\penalty0 103637, 2021.

\bibitem[Wornow et~al.(2024{\natexlab{b}})Wornow, Thapa, Steinberg, Fries, and Shah]{wornow2024ehrshot}
Michael Wornow, Rahul Thapa, Ethan Steinberg, Jason Fries, and Nigam Shah.
\newblock Ehrshot: An ehr benchmark for few-shot evaluation of foundation models.
\newblock \emph{Advances in Neural Information Processing Systems}, 36, 2024{\natexlab{b}}.

\bibitem[Wornow et~al.(2025)Wornow, Bedi, Hernandez, Steinberg, Fries, Re, Koyejo, and Shah]{wornowcontext}
Michael Wornow, Suhana Bedi, Miguel Angel~Fuentes Hernandez, Ethan Steinberg, Jason~Alan Fries, Christopher Re, Sanmi Koyejo, and Nigam Shah.
\newblock Context clues: Evaluating long context models for clinical prediction tasks on ehr data.
\newblock In \emph{The Thirteenth International Conference on Learning Representations}, 2025.

\bibitem[Jiang et~al.(2024)Jiang, Khan, Vasantha, and Haider]{jiang2024evidence}
Xiaorui Jiang, Kulsoom Khan, Sumithra~Thinakara Vasantha, and Sajjad Haider.
\newblock Evidence extraction for automated medical coding: preliminary evaluation.
\newblock In \emph{Proceedings of the 2024 8th International Conference on Natural Language Processing and Information Retrieval}, pages 18--23, 2024.

\bibitem[Pathak et~al.(2024)Pathak, Vald, Sermet, and Demir]{pathak2024utilizing}
Rudransh Pathak, Gabriel Vald, Yusuf Sermet, and Ibrahim Demir.
\newblock Utilizing large language models to predict icd-10 diagnosis codes from patient medical records.
\newblock In \emph{2024 IEEE MIT Undergraduate Research Technology Conference (URTC)}, pages 1--5. IEEE, 2024.

\bibitem[Lee and Lindsey(2024)]{lee2024can}
Simon~A Lee and Timothy Lindsey.
\newblock Can large language models abstract medical coded language?
\newblock \emph{arXiv preprint arXiv:2403.10822}, 2024.

\bibitem[Lee et~al.(2025{\natexlab{d}})Lee, Wu, and Chiang]{lee2025modern}
Simon~A Lee, Anthony Wu, and Jeffrey~N Chiang.
\newblock Clinical modernbert: An efficient and long context encoder for biomedical text.
\newblock \emph{arXiv preprint arXiv:2504.03964}, 2025{\natexlab{d}}.

\bibitem[Jo et~al.(2026)Jo, Kim, Son, and Kim]{jo2026clinical}
Hyunwoo Jo, Seon Kim, Hyunwoo Son, and Jongchan Kim.
\newblock Clinical text embeddings: A systematic review of methods, applications, and future directions.
\newblock \emph{International Journal of Medical Informatics}, page 106505, 2026.

\bibitem[Liang et~al.(2025)Liang, Guan, Zhang, Lian, Xu, and Lu]{liang2025aicoder}
Zhenpeng Liang, Hongjiao Guan, Weiyu Zhang, Ying Lian, Bing Xu, and Wenpeng Lu.
\newblock Aicoder: Exploring automated icd coding on chinese emrs with a multi-agent framework.
\newblock In \emph{2025 IEEE International Conference on Bioinformatics and Biomedicine (BIBM)}, pages 3823--3827. IEEE, 2025.

\bibitem[An et~al.(2026)An, Tang, Chen, and Guo]{an2026gate}
Ying An, Shiyu Tang, Xianlai Chen, and Lin Guo.
\newblock Gate: Graph and text exchange for zero-shot ecg classification with llm prompts.
\newblock \emph{IEEE Journal of Biomedical and Health Informatics}, 2026.

\bibitem[Zhu et~al.(2025)Zhu, Huang, Feng, Mu, Gu, Zhang, Dou, and Zhang]{zhu2025cp}
Yakun Zhu, Zhongzhen Huang, Qianhan Feng, Linjie Mu, Yannian Gu, Shaoting Zhang, Qi~Dou, and Xiaofan Zhang.
\newblock Cp-env: Evaluating large language models on clinical pathways in a controllable hospital environment.
\newblock \emph{arXiv preprint arXiv:2512.10206}, 2025.

\bibitem[Lazovi{\'c} et~al.(2025)Lazovi{\'c}, Madeira, Zdravevski, Silva, Coelho, and Pires]{lazovic2025analytical}
Kemal Lazovi{\'c}, Filipe Madeira, Eftim Zdravevski, Luis~Augusto Silva, Paulo~Jorge Coelho, and Ivan~Miguel Pires.
\newblock An analytical review of optimization techniques in information retrieval for enhanced decision support.
\newblock \emph{Decision Analytics Journal}, page 100657, 2025.

\bibitem[Zhi et~al.(2025)Zhi, Zhao, Wu, Zhao, and Zhu]{zhi2025reinventing}
Xiaoquan Zhi, Hongke Zhao, Likang Wu, Chuang Zhao, and Hengshu Zhu.
\newblock Reinventing clinical dialogue: Agentic paradigms for llm enabled healthcare communication.
\newblock \emph{arXiv preprint arXiv:2512.01453}, 2025.

\bibitem[Duong et~al.(2026)Duong, Le, Williams, Stewart, Zhao, Zitu, El~Naqa, Rollison, and Thieu]{duong2026oncopt}
Thanh Duong, Dung Le, Vonetta Williams, Sandra Stewart, Yayi Zhao, Muntasir Zitu, Issam El~Naqa, Dana Rollison, and Thanh Thieu.
\newblock Oncopt: long-context transformer models for in hospital tumor phenotype extraction from pathology reports.
\newblock \emph{npj Digital Medicine}, 2026.

\bibitem[Caron et~al.(2021)Caron, Touvron, Misra, J\'egou, Mairal, Bojanowski, and Joulin]{caron2021emerging}
Mathilde Caron, Hugo Touvron, Ishan Misra, Herv\'e J\'egou, Julien Mairal, Piotr Bojanowski, and Armand Joulin.
\newblock Emerging properties in self-supervised vision transformers.
\newblock In \emph{Proceedings of the International Conference on Computer Vision (ICCV)}, 2021.

\bibitem[Oquab et~al.(2023)Oquab, Darcet, Moutakanni, Vo, Szafraniec, Khalidov, Fernandez, Haziza, Massa, El-Nouby, et~al.]{oquab2023dinov2}
Maxime Oquab, Timoth{\'e}e Darcet, Th{\'e}o Moutakanni, Huy Vo, Marc Szafraniec, Vasil Khalidov, Pierre Fernandez, Daniel Haziza, Francisco Massa, Alaaeldin El-Nouby, et~al.
\newblock Dinov2: Learning robust visual features without supervision.
\newblock \emph{arXiv preprint arXiv:2304.07193}, 2023.

\bibitem[Radford et~al.(2021)Radford, Kim, Hallacy, Ramesh, Goh, Agarwal, Sastry, Askell, Mishkin, Clark, et~al.]{radford2021learning}
Alec Radford, Jong~Wook Kim, Chris Hallacy, Aditya Ramesh, Gabriel Goh, Sandhini Agarwal, Girish Sastry, Amanda Askell, Pamela Mishkin, Jack Clark, et~al.
\newblock Learning transferable visual models from natural language supervision.
\newblock In \emph{International conference on machine learning}, pages 8748--8763. PMLR, 2021.

\bibitem[Rombach et~al.(2021)Rombach, Blattmann, Lorenz, Esser, and Ommer]{rombach2021highresolution}
Robin Rombach, Andreas Blattmann, Dominik Lorenz, Patrick Esser, and Björn Ommer.
\newblock High-resolution image synthesis with latent diffusion models, 2021.

\bibitem[Saharia et~al.(2022)Saharia, Chan, Saxena, Li, Whang, Denton, Ghasemipour, Gontijo~Lopes, Karagol~Ayan, Salimans, et~al.]{saharia2022photorealistic}
Chitwan Saharia, William Chan, Saurabh Saxena, Lala Li, Jay Whang, Emily~L Denton, Kamyar Ghasemipour, Raphael Gontijo~Lopes, Burcu Karagol~Ayan, Tim Salimans, et~al.
\newblock Photorealistic text-to-image diffusion models with deep language understanding.
\newblock \emph{Advances in neural information processing systems}, 35:\penalty0 36479--36494, 2022.

\bibitem[Bertram et~al.(2024)Bertram, Fürnkranz, and Müller]{bertram2024contrastivelearningpreferencescontextual}
Timo Bertram, Johannes Fürnkranz, and Martin Müller.
\newblock Contrastive learning of preferences with a contextual infonce loss, 2024.
\newblock URL \url{https://arxiv.org/abs/2407.05898}.

\bibitem[Wang et~al.(2026)Wang, He, Zhang, Liu, Liu, Zeng, Qin, Li, Li, Yao, et~al.]{wang2026combating}
Zhaoqi Wang, Daqing He, Zijian Zhang, Ye~Liu, Jiamou Liu, Zhirui Zeng, Zhan Qin, Zhen Li, Xin Li, Hongwei Yao, et~al.
\newblock Combating knowledge corruption in agent systems: A byzantine-tolerant secure collaborative rag framework.
\newblock In \emph{Proceedings of the ACM Web Conference 2026}, pages 2661--2672, 2026.

\bibitem[Jain et~al.(2025)Jain, Coffee, He, Cron, Cochran, Mansilla-Gonzalez, Nadimpalli, Murad, and Osborne]{jain2025team}
Vijay~Raj Jain, Chris Coffee, Kaiwen He, Remy Cron, Micah~D Cochran, Luis Mansilla-Gonzalez, Akhil Nadimpalli, Danish Murad, and John~D Osborne.
\newblock Team uab at chemotherapy timelines 2025: Integrating encoders and large language models for chemotherapy timelines generation.
\newblock In \emph{Proceedings of the 7th Clinical Natural Language Processing Workshop}, pages 30--39, 2025.

\bibitem[Tian et~al.(2019)Tian, Krishnan, and Isola]{tian_2019_contrastic_distillation}
Yonglong Tian, Dilip Krishnan, and Phillip Isola.
\newblock Contrastive representation distillation.
\newblock \emph{arXiv}, 2019.
\newblock \doi{10.48550/arxiv.1910.10699}.
\newblock URL \url{https://arxiv.org/abs/1910.10699}.

\bibitem[Boguslav et~al.(2026)Boguslav, Kiehl, Kott, Strecker, Webb, Saklou, Ward, and Kirby]{boguslav2026fine}
Mayla~R Boguslav, Adam Kiehl, David Kott, George~Joseph Strecker, Tracy~L Webb, Nadia Saklou, Terri Ward, and Michael Kirby.
\newblock Fine-tuning foundational models to code diagnoses from veterinary health records.
\newblock \emph{PLOS Digital Health}, 5\penalty0 (2):\penalty0 e0001147, 2026.

\bibitem[T{\"u}rker et~al.(2025)T{\"u}rker, K{\i}z{\i}lo{\u{g}}lu, G{\"u}ng{\"o}r, and {\"U}sk{\"u}darl{\i}]{turker2025tabibert}
Melik{\c{s}}ah T{\"u}rker, A~Ebrar K{\i}z{\i}lo{\u{g}}lu, Onur G{\"u}ng{\"o}r, and Susan {\"U}sk{\"u}darl{\i}.
\newblock Tabibert: A large-scale modernbert foundation model and unified benchmarking framework for turkish.
\newblock \emph{arXiv preprint arXiv:2512.23065}, 2025.

\bibitem[Lindholz et~al.(2025)Lindholz, Burdenksi, Ruppel, Schulze-Weddige, Baumg{\"a}rtner, Schobert, Haack, Eminovic, Milnik, Hamm, et~al.]{lindholz2025comparing}
Maximilian Lindholz, Alina Burdenksi, Richard Ruppel, Sophia Schulze-Weddige, Georg~Lukas Baumg{\"a}rtner, Isabel Schobert, Anna-Maria Haack, Semil Eminovic, Annette Milnik, Charlie~Alexander Hamm, et~al.
\newblock Comparing large language models and text embedding models for automated classification of textual, semantic, and critical changes in radiology reports.
\newblock \emph{European Journal of Radiology}, page 112316, 2025.

\bibitem[Zhang and Li(2025)]{zhang2025chronoformer}
Yuanyun Zhang and Shi Li.
\newblock Chronoformer: Time-aware transformer architectures for structured clinical event modeling.
\newblock \emph{arXiv preprint arXiv:2504.07373}, 2025.

\bibitem[Rahman et~al.(2026{\natexlab{b}})Rahman, YongZhong, and Bin]{rahman2026graph}
Muhammad Rahman, Cao YongZhong, and Li~Bin.
\newblock Graph attention network-based multimodal approach for lung diseases classification.
\newblock \emph{Scientific Reports}, 2026{\natexlab{b}}.

\bibitem[Cui et~al.(2026)Cui, Swingle, Joshi, Nair, and Leahy]{cui2026predicting}
Wenhui Cui, Nicholas Swingle, Anand~A Joshi, Dileep Nair, and Richard~M Leahy.
\newblock Predicting post-traumatic epilepsy from clinical records using large language model embeddings.
\newblock \emph{arXiv preprint arXiv:2604.14547}, 2026.

\bibitem[Dao(2024)]{dao2023flashattention2}
Tri Dao.
\newblock Flash{A}ttention-2: Faster attention with better parallelism and work partitioning.
\newblock In \emph{International Conference on Learning Representations (ICLR)}, 2024.

\bibitem[Liu et~al.(2021)Liu, Lin, Cao, Hu, Wei, Zhang, Lin, and Guo]{liu2021swin}
Ze~Liu, Yutong Lin, Yue Cao, Han Hu, Yixuan Wei, Zheng Zhang, Stephen Lin, and Baining Guo.
\newblock Swin transformer: Hierarchical vision transformer using shifted windows.
\newblock In \emph{Proceedings of the IEEE/CVF international conference on computer vision}, pages 10012--10022, 2021.

\bibitem[Dosovitskiy et~al.(2021)Dosovitskiy, Beyer, Kolesnikov, Weissenborn, Zhai, Unterthiner, Dehghani, Minderer, Heigold, Gelly, Uszkoreit, and Houlsby]{dosovitskiy2021an}
Alexey Dosovitskiy, Lucas Beyer, Alexander Kolesnikov, Dirk Weissenborn, Xiaohua Zhai, Thomas Unterthiner, Mostafa Dehghani, Matthias Minderer, Georg Heigold, Sylvain Gelly, Jakob Uszkoreit, and Neil Houlsby.
\newblock An image is worth 16x16 words: Transformers for image recognition at scale.
\newblock In \emph{International Conference on Learning Representations}, 2021.
\newblock URL \url{https://openreview.net/forum?id=YicbFdNTTy}.

\bibitem[Wu et~al.(2023)Wu, Xiao, Liu, Yin, Pechenizkiy, Mocanu, Van~Keulen, and Mocanu]{wu2023e2enet}
Boqian Wu, Qiao Xiao, Shiwei Liu, Lu~Yin, Mykola Pechenizkiy, Decebal~Constantin Mocanu, Maurice Van~Keulen, and Elena Mocanu.
\newblock E2enet: Dynamic sparse feature fusion for accurate and efficient 3d medical image segmentation.
\newblock \emph{NeurIPS}, 2023.

\bibitem[Lai et~al.(2024)Lai, Jiang, Yao, Wang, He, Tao, Wei, Lv, and Zhou]{lai2024e3d}
Haoran Lai, Zihang Jiang, Qingsong Yao, Rongsheng Wang, Zhiyang He, Xiaodong Tao, Wei Wei, Weifu Lv, and S~Kevin Zhou.
\newblock E3d-gpt: Enhanced 3d visual foundation for medical vision-language model.
\newblock \emph{arXiv preprint arXiv:2410.14200}, 2024.

\bibitem[Shaker et~al.(2024)Shaker, Maaz, Rasheed, Khan, Yang, and Khan]{shaker2024unetr++}
Abdelrahman~M Shaker, Muhammad Maaz, Hanoona Rasheed, Salman Khan, Ming-Hsuan Yang, and Fahad~Shahbaz Khan.
\newblock Unetr++: delving into efficient and accurate 3d medical image segmentation.
\newblock \emph{IEEE Transactions on Medical Imaging}, 2024.

\bibitem[Xing et~al.(2024)Xing, Ye, Yang, Liu, and Zhu]{xing2024segmamba}
Zhaohu Xing, Tian Ye, Yijun Yang, Guang Liu, and Lei Zhu.
\newblock Segmamba: Long-range sequential modeling mamba for 3d medical image segmentation.
\newblock In \emph{International Conference on Medical Image Computing and Computer-Assisted Intervention}, pages 578--588. Springer, 2024.

\bibitem[Liu et~al.(2024)Liu, Xu, Woicik, Shapiro, Blazes, Wu, Lee, Lee, and Wang]{liu2024octcube}
Zixuan Liu, Hanwen Xu, Addie Woicik, Linda~G Shapiro, Marian Blazes, Yue Wu, Cecilia~S Lee, Aaron~Y Lee, and Sheng Wang.
\newblock Octcube: a 3d foundation model for optical coherence tomography that improves cross-dataset, cross-disease, cross-device and cross-modality analysis.
\newblock \emph{arXiv preprint arXiv:2408.11227}, 2024.

\bibitem[Choy et~al.(2019)Choy, Gwak, and Savarese]{choy20194d}
Christopher Choy, JunYoung Gwak, and Silvio Savarese.
\newblock 4d spatio-temporal convnets: Minkowski convolutional neural networks.
\newblock In \emph{Proceedings of the IEEE Conference on Computer Vision and Pattern Recognition}, pages 3075--3084, 2019.

\bibitem[Wang et~al.(2023)Wang, Wu, Luo, Liu, Li, and Zhang]{wang2023mis}
Guotai Wang, Jianghao Wu, Xiangde Luo, Xinglong Liu, Kang Li, and Shaoting Zhang.
\newblock Mis-fm: 3d medical image segmentation using foundation models pretrained on a large-scale unannotated dataset.
\newblock \emph{arXiv preprint arXiv:2306.16925}, 2023.

\bibitem[Yakdan et~al.(2026)Yakdan, Warner, Ghogawala, Ray, Bydon, Steinmetz, Griffey, Foraker, Wilcox, Lu, et~al.]{yakdan2026clinically}
Salim Yakdan, Ben Warner, Zoher Ghogawala, Wilson~Z Ray, Mohamad Bydon, Michael~P Steinmetz, Richard~T Griffey, Randi Foraker, Adam Wilcox, Chenyang Lu, et~al.
\newblock Clinically-guided models or foundation models? predicting cervical spondylotic myelopathy from electronic health records.
\newblock \emph{npj Digital Medicine}, 2026.

\bibitem[von Arnim et~al.(2026)von Arnim, Kohler, Hegselmann, Marschollek, Eils, and Wild]{von2026rethinking}
Georg von Arnim, Severin Kohler, Stefan Hegselmann, Michael Marschollek, Roland Eils, and Benjamin Wild.
\newblock Rethinking healthcare data interoperability in the age of large language models.
\newblock \emph{Med}, 2026.

\bibitem[Contreras et~al.(2025)Contreras, Kapoor, Zhang, Davidson, Ren, Guan, Ozrazgat-Baslanti, Sena, Nerella, Bihorac, et~al.]{contreras2025large}
Miguel Contreras, Sumit Kapoor, Jiaqing Zhang, Andrea Davidson, Yuanfang Ren, Ziyuan Guan, Tezcan Ozrazgat-Baslanti, Jessica Sena, Subhash Nerella, Azra Bihorac, et~al.
\newblock A large language model for delirium prediction in the intensive care unit using structured electronic health records.
\newblock \emph{Scientific Reports}, 15\penalty0 (1):\penalty0 38890, 2025.

\bibitem[Li et~al.(2024)Li, Qu, Chen, Bassi, Shi, Lai, Yu, Xue, Chen, Lin, et~al.]{li2024abdomenatlas}
Wenxuan Li, Chongyu Qu, Xiaoxi Chen, Pedro~RAS Bassi, Yijia Shi, Yuxiang Lai, Qian Yu, Huimin Xue, Yixiong Chen, Xiaorui Lin, et~al.
\newblock Abdomenatlas: A large-scale, detailed-annotated, \& multi-center dataset for efficient transfer learning and open algorithmic benchmarking.
\newblock \emph{Medical Image Analysis}, 97:\penalty0 103285, 2024.

\bibitem[Tizhoosh(2025)]{tizhoosh2025beyond}
Hamid~R Tizhoosh.
\newblock Beyond the failures: Rethinking foundation models in pathology.
\newblock \emph{arXiv preprint arXiv:2510.23807}, 2025.

\end{thebibliography}


\end{document}